\documentclass[preprint,showpacs,preprintnumbers,amsmath,amssymb,nofootinbib]{revtex4-2}

\usepackage{graphicx}
\usepackage{color}
\usepackage{comment}
\usepackage[normalem]{ulem}
\usepackage{bm}

\begin{document}

\title{Impact of charge-density-wave pattern on the superconducting gap in V-based kagome superconductors} 

\author{T.~Nagashima$^{1,\S}$}
\author{K.~Ishihara$^{1,\S}$}\email{k.ishihara@edu.k.u-tokyo.ac.jp}
\author{Y.~Yamakawa$^2$}
\author{F.~Chen$^1$}
\author{K.~Imamura$^1$}
\author{M.~Roppongi$^1$}
\author{R.~Grasset$^3$}
\author{M.~Konczykowski$^3$}
\author{B.~R.~Ortiz$^4$}
\author{A.~C.~Salinas$^5$}
\author{S.~D.~Wilson$^5$}
\author{R.~Tazai$^6$}
\author{H.~Kontani$^2$}
\author{K.~Hashimoto$^1$}
\author{T.~Shibauchi$^1$}\email{shibauchi@k.u-tokyo.ac.jp}
\affiliation{$^1$Department of Advanced Materials Science, University of Tokyo, Kashiwa, Chiba 277-8561, Japan\\
$^2$Department of Physics, Nagoya University, Nagoya 464-8602, Japan\\
$^3$Laboratoire des Solides Irradi{\' e}s, CEA/DRF/IRAMIS, Ecole Polytechnique, CNRS, Institut Polytechnique de Paris, F-91128 Palaiseau, France.\\
$^4$Materials Science and Technology Division, Oak Ridge National Laboratory, Oak Ridge, TN, USA\\
$^5$Materials Department, University of California Santa Barbara, Santa Barbara, CA 93106, USA.\\
$^6$Yukawa Institute for Theoretical Physics, Kyoto University, Kyoto 606-8502, Japan\\
$^\S$\rm These authors contributed equally to this work.}

\date{\today}
\maketitle


{\bf
Kagome metals $\bm{A}$V$\bm{_3}$Sb$\bm{_5}$ ($\bm{A}=$ K, Rb, Cs) provide a compelling platform to explore the interplay between superconductivity (SC) and charge-density-wave (CDW) orders. While distinct CDW orders have been identified in K/RbV$\bm{_3}$Sb$\bm{_5}$ versus CsV$\bm{_3}$Sb$\bm{_5}$, their influence on the SC order parameter remains unresolved. Here, we investigate low-energy quasiparticle excitations in $\bm{A}$V$\bm{_3}$Sb$\bm{_5}$, uncovering a striking difference in SC gap anisotropy: K/RbV$\bm{_3}$Sb$\bm{_5}$ exhibit fully gapped, nearly isotropic $\bm{s}$-wave states, in contrast to the strongly anisotropic SC gap in CsV$\bm{_3}$Sb$\bm{_5}$. Contrary to previous vortex-state studies suggesting nodal SC in K/RbV$\bm{_3}$Sb$\bm{_5}$, our Meissner-state measurements in high-quality crystals demonstrate fully gapped states with reduced anisotropy compared to CsV$_3$Sb$_5$. Impurity scattering introduced via electron irradiation in K/RbV$_3$Sb$_5$ has a minimal impact on low-energy excitations, and it induces an increase in the SC transition temperature $\bm{T_{\rm c}}$, consistent with more isotropic $\bm{s}$-wave SC competing with CDW order. 
Our theoretical analysis attributes the observed SC gap anisotropy differences to distinct CDW modulation patterns: the star-of-David structure unique to CsV$\bm{_3}$Sb$\bm{_5}$ preserves van Hove singularities near the Fermi level, promoting anisotropic $\bm{s}$-wave SC with enhanced $\bm{T_{\rm c}}$ via bond-order fluctuations. 
These findings establish a systematic framework for understanding the interplay between SC and CDW orders in $\bm{A}$V$\bm{_3}$Sb$\bm{_5}$, driven by electron correlations.}
\clearpage


Kagome lattices, two-dimensional networks of corner-sharing triangles, offer fertile playgrounds for various quantum phases due to their intrinsic geometrical frustrations and unique band structures, including flat bands, Dirac cones at $K$ points, and van Hove singularities (vHSs) at $M$ points. Especially when the Fermi level is located near vHSs, the divergent density of states (DOS), good nesting between $M$ points, and characteristic orbital nature of vHSs give rise to a wide variety of emergent exotic orders such as spin-density-wave, CDW, loop-current orders, and unconventional SC\,\cite{Wang2013,Kiesel2013,Wu2021,Tazai2022,Tazai2023}. Furthermore, recent efforts have been devoted to exploring novel emergent physics originating from an interplay between fascinating orders in kagome materials.

\begin{figure*}[tbp]
    \includegraphics[width=0.7\linewidth]{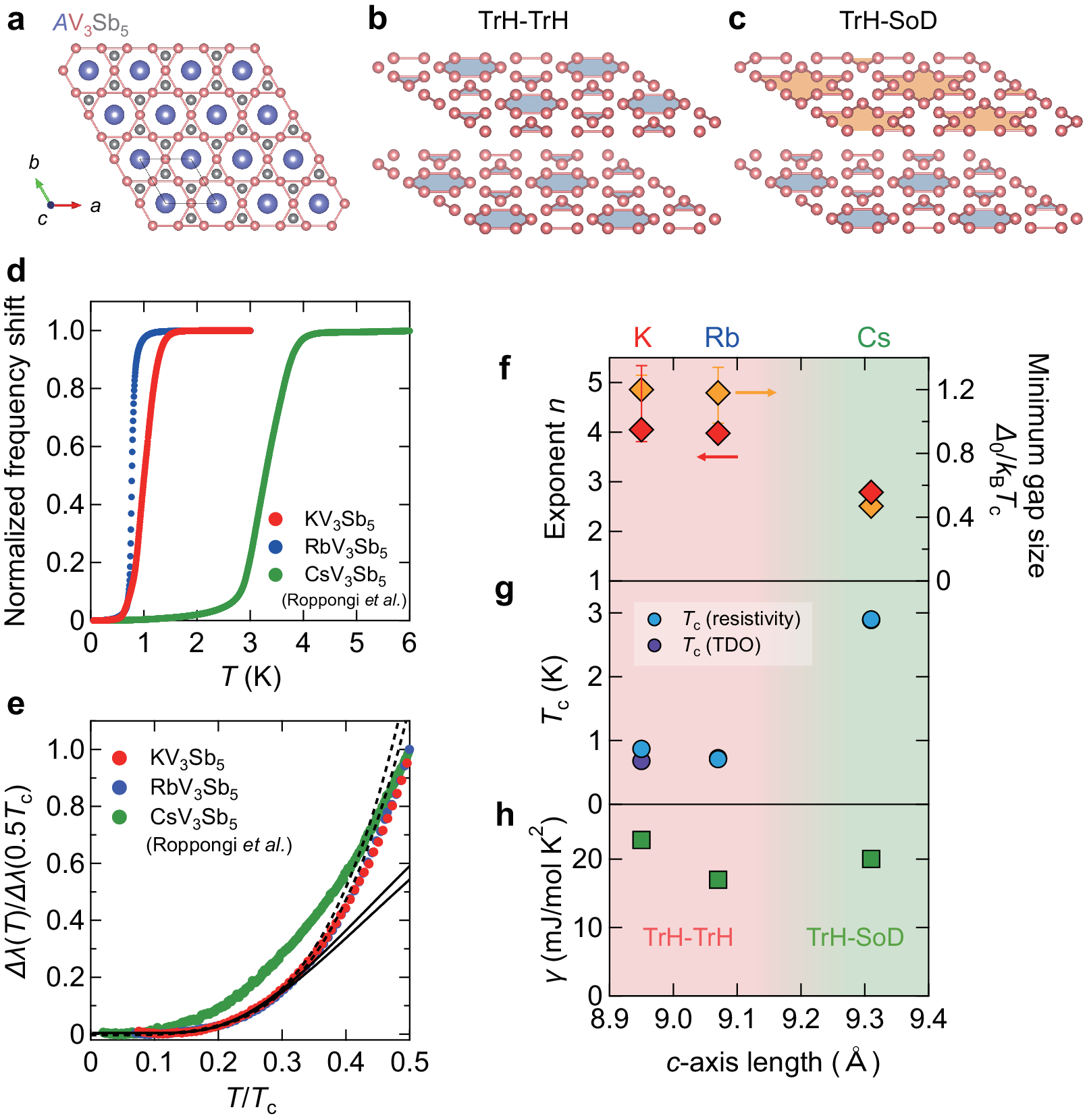}
    \caption{{\bf Crystal structure, CDW modulations, and penetration depth measurements in \textit{A}V$_3$Sb$_5$ (\textit{A} = K, Rb, Cs).} {\bf a}, Crystal structure of \textit{A}V$_3$Sb$_5$. {\bf b}, {\bf c}, The three-dimensional BO structure in \textit{A}V$_3$Sb$_5$: TrH order with $\pi$ phase shift between neighboring planes ({\bf b}), and a mixture of TrH and SoD orders ({\bf c}). {\bf d}, Temperature dependence of the normalized frequency shift of the TDO for the pristine samples of \textit{A}V$_3$Sb$_5$. The data for CsV$_3$Sb$_5$ are taken from Ref.\,\cite{Roppongi2023}. {\bf e}, Temperature dependence of the penetration depth below 0.5$T_{\rm c}$ normalized by the value at 0.5$T_{\rm c}$ for the pristine samples of \textit{A}V$_3$Sb$_5$. The black dashed lines and black solid lines represent fitting curves with power-law and fully-gapped models, respectively. {\bf f}, Exponent $n$ obtained from the fitting with power-law below 0.3$T_{\rm c}$ (red diamonds), minimum gap size $\Delta_0/k_{\rm{B}}T_{\rm c}$ obtained from the fully-gapped fitting (orange diamonds) as a function of $c$-axis length. The error bar is estimated by varying the fitting range $0.25\leq T_{\rm max}/T_{\rm c}\leq 0.35$ (see Supplementary Information). {\bf g}, $T_{\rm c}$ determined from the resistivity measurements by linearly extrapolating the resistivity in the transition region to zero (light blue circles), and from the TDO measurements, where superfluid density becomes finite (purple circles). {\bf h}, Sommerfeld coefficient $\gamma$ obtained from the specific heat measurements\,\cite{Yin2021, Ortiz2021, Duan2021}.}
    \label{F1}
\end{figure*}

Recently discovered kagome metals $A$V$_3$Sb$_5$ ($A =$ K, Rb, Cs), consisting of perfect kagome networks of V atoms (Fig.\,\ref{F1}{\bf a}), have attracted considerable attention due to their band structure with the Fermi energy situated near vHSs at $M$ points\,\cite{Wilson2024}. In all alkali compounds, CDW and SC orders coexist with the CDW transition temperature $T_{\rm CDW}$ of 78, 104, and 94\,K and the SC transition temperature $T_{\rm c}$ of 0.9, 0.8, and 2.5\,K for $A=$ K, Rb, and Cs, respectively\,\cite{Ortiz2020,Ortiz2021,Yin2021}. A strong interplay between the SC and CDW orders is implied from the pressure-temperature ($P$-$T$) phase diagram, where $T_{\rm c}$ shows a peak structure around the CDW endpoint\,\cite{Du2021,Wang2021,Chen2021,Yu2021}. The CDW order is expected to be quite exotic, with time-reversal symmetry (TRS)\,\cite{Yang2020,Jiang2021,Guo2022,Mielke2022, Xu2022, Asaba2024} and rotational symmetry\,\cite{Zhao2021, Xiang2021, Li2022, Nie2022, Xu2022, Li2023, Asaba2024} breaking, as well as translational symmetry breaking. The coexistence of bond order (BO) and loop-current order, where the former corresponds to a real CDW order that preserves TRS and the latter corresponds to an imaginary CDW order that breaks TRS, has been theoretically proposed to understand the multiple symmetry breaking in kagome metals\,\cite{Tazai2023,Tazai2024}.

Although the overall band structure is similar in all alkali compounds, some differences in electronic states between K/RbV$_3$Sb$_5$ and CsV$_3$Sb$_5$ have been recently pointed out\,\cite{Wilson2024}. For example, although the BO within the kagome planes is established as the 2$\times$2 star of David (SoD) or tri-hexagonal (TrH) structure by scanning tunneling microscopy (STM) measurements\,\cite{Jiang2021,Shumiya2021,Zhao2021}, the out-of-plane modulation pattern of the BO varies between K/RbV$_3$Sb$_5$ and CsV$_3$Sb$_5$. Specifically, 2$\times$2$\times$2 staggered TrH order with $\pi$ phase shift (Fig.\,\ref{F1}{\bf b}) is expected in K/RbV$_3$Sb$_5$\,\cite{Kang2023,Kautzsch2023,Kato2022,Frassineti2023}, and 2$\times$2$\times$2 and 2$\times$2$\times$4 more complex patterns with coexisting SoD and TrH orders (Fig.\,\ref{F1}{\bf c}) are discussed in CsV$_3$Sb$_5$ \,\cite{Kang2023,Kautzsch2023,Ortiz2021PRX,Li2022PRR,Hu2022}. Indeed, differences in the momentum-dependent CDW gap and structural deformations associated with the CDW transition have been detected by angle-resolved photoemission spectroscopy (ARPES) and x-ray diffraction measurements between K/RbV$_3$Sb$_5$ and CsV$_3$Sb$_5$\,\cite{Kang2023,Kautzsch2023}. As the origin of the distinct CDW patterns, the involvement of vHSs having different orbital characters below the Fermi energy has been theoretically discussed\,\cite{Wilson2024}. It is noteworthy that, in the $P$-$T$ and hole-doping $A$V$_3$Sb$_{5-x}$Sn$_x$ phase diagrams, the SC phase shows a single peak at the CDW endpoint in $A$ = K and Rb\,\cite{Du2021, Wang2021, Oey2022_KRb}, while in $A$ = Cs it shows double peaks: one within the CDW phase and the other at the CDW endpoint\,\cite{Chen2021,Yu2021,Oey2022_Cs}. These distinct peak structures are likely associated with the different CDW patterns between K/RbV$_3$Sb$_5$ and CsV$_3$Sb$_5$, and subsequent changes in the CDW state under pressure or hole-doping in CsV$_3$Sb$_5$\,\cite{Kang2023,Zheng2022}.

As for the SC symmetry, theories based on $A$V$_3$Sb$_5$ have proposed the spin-triplet $p$- and $f$-wave, spin-singlet chiral $d$-wave, and nodal or nodeless $s$-wave states\,\cite{Wang2013,Kiesel2013,Wu2021,Tazai2022}. From the experimental point of view, some of the authors have reported systematic penetration depth measurements by changing the impurity concentrations in CsV$_3$Sb$_5$, indicating an anisotropic nodeless $s$-wave pairing without sign change in the gap function\,\cite{Roppongi2023}, in line with nuclear quadrupole resonance\,\cite{Mu2021}, muon spin relaxation ($\mu$SR)\,\cite{Gupta2022}, STM\,\cite{Xu2021}, and similar penetration depth measurements\,\cite{Duan2021,Grant2024}. Especially a recent ARPES study on CsV$_3$Sb$_5$ reported the coexistence of anisotropic and isotropic SC gaps in the momentum space\,\cite{Mine2024}, which confirms the validity of two-band model used in Ref.\,\cite{Roppongi2023}. 
On the other hand, the gap structure in K/RbV$_3$Sb$_5$ has been rarely investigated except for a $\mu$SR study suggesting a nodal gap structure from a linear-in-$T$ behavior of the superfluid density at low temperatures\,\cite{Guguchia2023}. Thus, a systematic understanding of the SC gap structure in the $A$V$_3$Sb$_5$ family and the relationship between the SC and CDW orders remain elusive.


The interplay between the SC and CDW orders in $A$V$_3$Sb$_5$ has been studied through the phase diagrams under pressure or chemical substitutions\,\cite{Du2021,Wang2021,Chen2021,Yu2021,Oey2022_KRb,Oey2022_Cs}. However, these approaches alter the lattice constants, resulting in a change in the band structures, which may mask a pure interrelation between the SC and CDW orders. In this study, we instead focus on impurity effects on the SC and CDW states, which have been applied to study the relationship between SC and CDW orders in cuprates and transition metal dichalcogenides (TMDs) without affecting the crystal and electronic structure\,\cite{Cho2018,Leroux2019}. Furthermore, impurity effects are quite useful to investigate the SC symmetry because the conventional SC is robust against nonmagnetic impurities, while unconventional SC states with strong gap anisotropy can be suppressed by disorder\,\cite{Roppongi2023}. Moreover, nonmagnetic impurity scatterings induce additional low-energy Andreev bound states when the gap function has a sign change, which can be detected by low-energy quasiparticle excitation measurements\,\cite{Mizukami2014,Nagashima2024}. Therefore, the combination of measurements of quasiparticle excitations and their impurity effects is a phase-sensitive bulk probe of the gap symmetry, which can distinguish, for example, fully-gapped $s$-wave and chiral $d$-wave pairings. In this study, we performed magnetic penetration depth measurements by the tunnel diode oscillator (TDO) technique combined with a systematic control of impurity concentrations using high-energy electron irradiation in K/RbV$_3$Sb$_5$ to provide a comprehensive understanding of the SC symmetry and the interplay between the SC and CDW orders in the $A$V$_3$Sb$_5$ family.


First, we discuss the SC gap structure of $A$V$_3$Sb$_5$ from the temperature dependence of the penetration depth in pristine samples. Figure\,\ref{F1}{\bf d} represents the temperature dependence of the normalized frequency shift in the TDO for all alkali compounds. We note that the data of CsV$_3$Sb$_5$ are taken from the previous study\,\cite{Roppongi2023}. A slightly broad SC transition in Fig.\,\ref{F1}{\bf d} may be related to the SC phase fluctuations or a short skin depth. Figure\,\ref{F1}{\bf g} summarizes $T_{\rm c}$ in $A$V$_3$Sb$_5$, determined from the temperature dependence of the normalized superfluid density $\rho_s(T)$, which is derived from the change in the penetration depth $\Delta \lambda (T)$, as in the case of CsV$_3$Sb$_5$\,\cite{Roppongi2023}. Figure\,\ref{F1}{\bf e} shows $\Delta \lambda (T)$ normalized by the 0.5$T_{\rm c}$ value in the pristine $A$V$_3$Sb$_5$ samples. As clearly seen, CsV$_3$Sb$_5$ shows larger excitations at low temperatures compared to K/RbV$_3$Sb$_5$. To evaluate the temperature dependence of these results more quantitatively, we fitted the data below $T/T_{\rm c}<0.3$ with a power-law function, $\Delta \lambda (T) \propto T^n$, where $n\leq 2$ ($n>2$) implies a nodal (fully gapped) gap structure. The dashed lines in Fig.\,\ref{F1}{\bf e} represent the fitting curves, and the obtained exponent $n$ values are summarized in Fig.\,\ref{F1}{\bf f}. The result of $n>2$ for all $A$V$_3$Sb$_5$ compounds suggests that the fully-gapped SC is universally realized in the $A$V$_3$Sb$_5$ family. It is worth noting that this result contradicts the $\mu$SR study suggesting a nodal gap structure in K/RbV$_3$Sb$_5$\,\cite{Guguchia2023}. The discrepancy between our results and the previous report is discussed in Supplementary Information. More importantly, although all alkali compounds exhibit fully gapped behaviors, $\Delta \lambda (T)$ substantially differs between K/RbV$_3$Sb$_5$ and CsV$_3$Sb$_5$. To clarify the differences, we fitted $\Delta \lambda (T)$ below 0.3$T_{\rm{c}}$ with a fully gapped model, $\Delta\lambda (T) \propto T^{-1/2} \exp(-\Delta_0/k_{\rm B}T)$, where $\Delta_0$ is the gap minima in the momentum space and $k_{\rm B}$ is the Boltzmann constant. The obtained $\Delta_0/k_{\rm B}T_{\rm c}$ values are also depicted in Fig.\,\ref{F1}{\bf f}, signifying that $\Delta_0/k_{\rm B}T_{\rm c}$ in K/RbV$_3$Sb$_5$ is clearly larger than that in CsV$_3$Sb$_5$. Considering that an anisotropic gap structure makes the $\Delta_0$ value smaller, our results of the pristine $A$V$_3$Sb$_5$ samples indicate that the gap structure in K/RbV$_3$Sb$_5$ is more isotropic than in CsV$_3$Sb$_5$.

\begin{figure}[tbp]
    \includegraphics[width=0.9\linewidth]{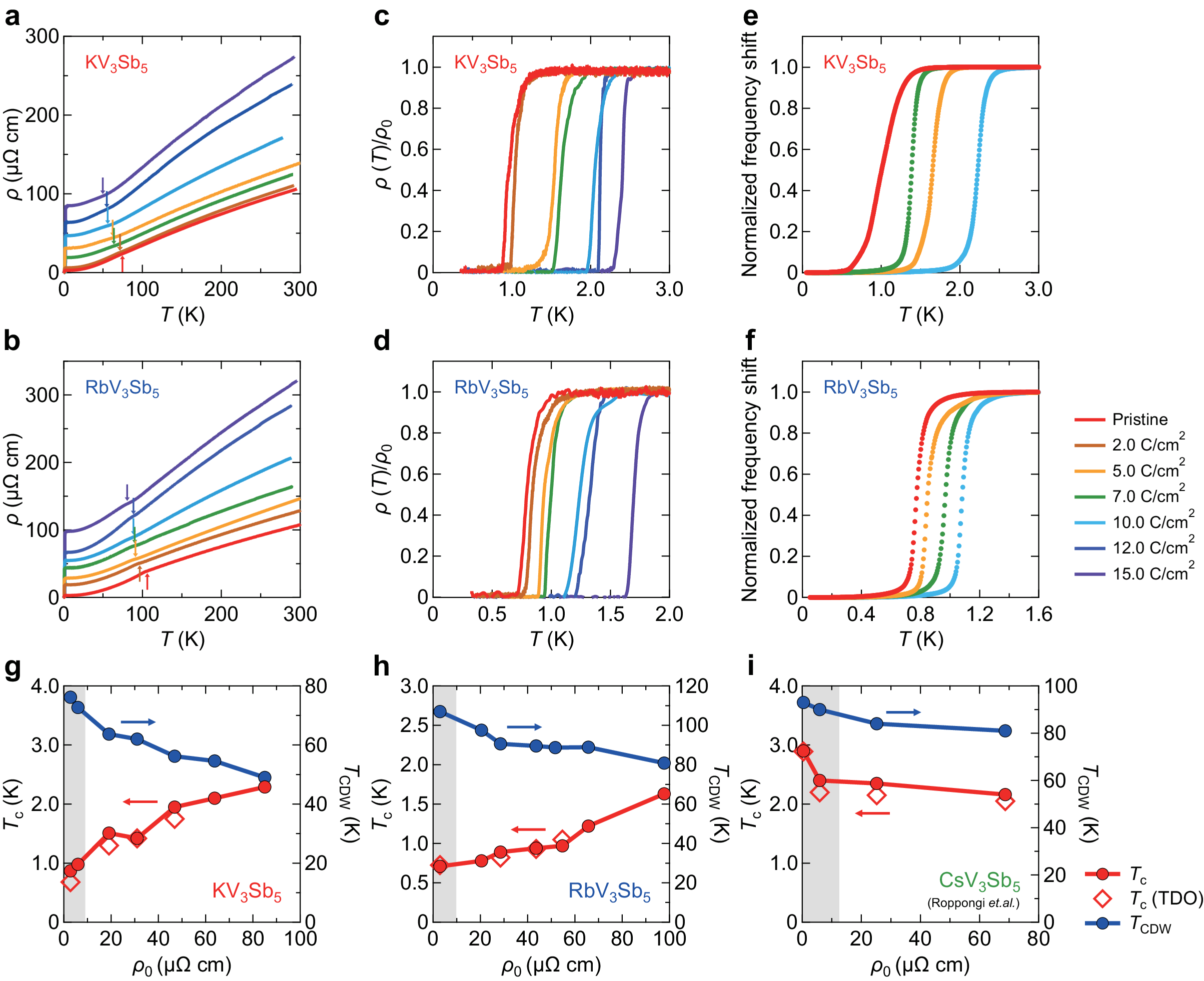}
    \caption{{\bf Impurity effects on the transition temperatures in \textit{A}V$_3$Sb$_5$ (\textit{A} = K, Rb, Cs).} {\bf a}, {\bf b}, Temperature dependence of the electrical resistivity $\rho(T)$ for various electron irradiation doses in KV$_3$Sb$_5$ ({\bf a}) and RbV$_3$Sb$_5$ ({\bf b}). Arrows indicate the CDW transition temperatures. {\bf c}, {\bf d},  $\rho(T)$ normalized by the residual resistivity $\rho_0$ around the superconducting transition in pristine and irradiated KV$_3$Sb$_5$ ({\bf c}) and RbV$_3$Sb$_5$ ({\bf d}). {\bf e}, {\bf f}, Temperature dependence of the normalized frequency shift in the TDO measurements for various irradiation doses in KV$_3$Sb$_5$ ({\bf e}) and RbV$_3$Sb$_5$ ({\bf f}). {\bf g}-{\bf i}, Changes in $T_{\rm{c}}$ and $T_{\rm{CDW}}$ as a function of $\rho_0$ in KV$_3$Sb$_5$ ({\bf g}), RbV$_3$Sb$_5$ ({\bf h}), and CsV$_3$Sb$_5$ ({\bf i}). $T_{\rm c}$ is obtained from the resistivity (filled red circles) and TDO measurements (open red diamonds).  $T_{\rm{CDW}}$ (filled brue circles) is determined from the resistivity measurements as a temperature at which $d\rho /dT$ exhibits dip (see Supplemental Information). The gray region represents the parameter space where SC with sign-changing order parameters is expected to survive based on the Abrikosov-Gor'kov (AG) therory. This region is determined using the pair-breaking parameter $g=\hbar/\tau_{\rm imp}k_{\rm B}T_{\rm c0}$, where $\tau_{\rm imp}=\mu_0 \lambda^2(0)/\rho_0$ and $T_{\rm c0}$ is the SC transition temperature of the pristine sample. We estimate $\lambda(0)$ from the $\mu$SR measurements\,\cite{Guguchia2023, Gupta2022}.}
    \label{F2}
\end{figure}

Next, we focus on the impurity effects on $T_{\rm c}$ and $T_{\rm CDW}$. Figures\,\ref{F2}{\bf a}, {\bf b} show the temperature dependence of electrical resistivity $\rho(T)$ in pristine and electron-irradiated samples of KV$_3$Sb$_5$ and RbV$_3$Sb$_5$, respectively. The resistivity progressively increases with irradiation, suggesting the successive introduction of homogeneous impurities by electron irradiation. Defining $T_{\rm CDW}$ from an anomaly in ${\rm d}\rho(T)/{\rm d}T$ as depicted by arrows in Figs.\,S4{\bf a}, {\bf b}, a clear decrease in $T_{\rm CDW}$ with increasing impurity scattering can be observed in both the K and Rb cases, reminiscent of the case in CsV$_3$Sb$_5$. Focusing on the low-$T$ behavior of $\rho(T)$ in Figs.\,\ref{F2}{\bf c}, {\bf d}, on the other hand, $T_{\rm c}$ shows a clear increase with irradiation in contrast to $T_{\rm CDW}$, which is a completely opposite behavior to the CsV$_3$Sb$_5$ case where $T_{\rm c}$ is suppressed by irradiation. The increase in $T_{\rm c}$ with disorder is also confirmed by the TDO measurements, as shown in Figs.\,\ref{F2}{\bf e}, {\bf f}. Figures\,\ref{F2}{\bf g}-{\bf i} summarize $T_{\rm c}$ and $T_{\rm CDW}$ as a function of residual resistivity $\rho_0$ for $A$ = K, Rb, and Cs, respectively. For all compounds, SC survives beyond the $\rho_0$ range where the SC with sign changing in gap functions can survive by the Abrikosov-Gor'kov theory (gray regions in Figs.\,\ref{F2}{\bf g}-{\bf i}). This result rules out the SC states with symmetry-protected sign changing gap, such as chiral $d$-wave, $p$-wave, and $f$-wave pairing states. From the perspective of the relationship between SC and CDW orders, the opposite trend with decreasing $T_{\rm CDW}$ and increasing $T_{\rm c}$ indicates strong competition between SC and CDW states. A similar competition has been reported in some cuprates and TMDs\,\cite{Cho2018,Leroux2019}. However, the origin of the opposite behavior between K/RbV$_3$Sb$_5$ and CsV$_3$Sb$_5$ even with the similar electronic structure is nontrivial, which will be discussed later.

\begin{figure}[tbp]
    \includegraphics[width=\linewidth]{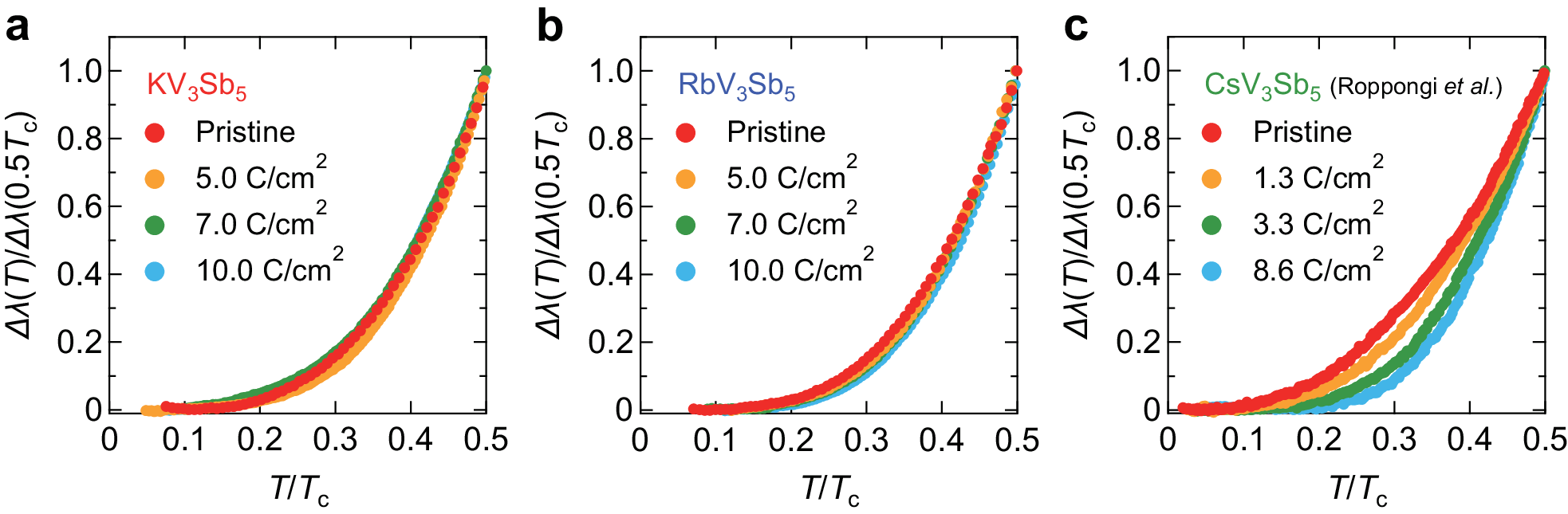}
    \caption{{\bf Impurity effects on the temperature dependence of the penetration depth.}  {\bf a}-{\bf c}, Temperature dependence of the penetration depth below 0.5$T_{\rm{c}}$ normalized by the value at 0.5$T_{\rm c}$ for various irradiation doses in KV$_3$Sb$_5$ ({\bf a}), RbV$_3$Sb$_5$ ({\bf b}), and CsV$_3$Sb$_5$ ({\bf c}). }
    \label{F3}
\end{figure}

The impurity effects on the gap structure can be discussed from the evolution in the low-$T$ behavior of $\Delta \lambda (T)$ against electron irradiation, as shown in Fig.\,\ref{F3}. As reported in Ref.\,\cite{Roppongi2023}, the gap structure in CsV$_3$Sb$_5$ significantly changes with irradiation, indicating that the anisotropic gap structure in the pristine sample becomes more isotropic by impurity scattering via the gap averaging effect in the momentum space. This can be seen in $\Delta \lambda (T)$ shown in Fig.\,\ref{F3}{\bf c}, where flattening behaviors are observed at low temperatures in irradiated samples. In K/RbV$_3$Sb$_5$, on the other hand, $\Delta \lambda (T)$ remains intact against impurity scattering (Figs.\,\ref{F3}{\bf a}, {\bf b}), which is consistent with the more isotropic gap structure. This is because when the gap structure of the pristine sample is isotropic, the gap averaging effect can no longer play a role. Furthermore, since $\Delta \lambda (T) \propto T^2$ is expected to be caused by the impurity-induced low-energy Andreev bound states in the dirty SC state with sign-changing gap function, the observed robust behavior of $\Delta \lambda (T)$ with exponential $T$ dependence against impurities provides strong evidence for an isotropic sign-preserving SC gap structure in both KV$_3$Sb$_5$ and RbV$_3$Sb$_5$.


The opposite trend of $T_{\rm c}$ against irradiation between K/RbV$_3$Sb$_5$ and CsV$_3$Sb$_5$ can be explained as follows based on the observed SC gap structure. Whether $T_{\rm c}$ increases or decreases with impurities is determined by the balance of two contributions; one is the relationship between SC and coexisting electronic orders, and the other is the gap-averaging effect in the momentum space. Here, the former increases $T_{\rm c}$ when the order competing with SC is suppressed, and the latter decreases $T_{\rm c}$. From $\Delta \lambda (T)$ shown in Fig.\,\ref{F3}, the gap-averaging effect is substantial in CsV$_3$Sb$_5$, while that in K/RbV$_3$Sb$_5$ is relatively weak. Therefore, although the latter contribution, which reduces $T_{\rm c}$, plays a significant role in CsV$_3$Sb$_5$, we can purely detect the former contribution originating from the intense competition between SC and CDW orders in K/RbV$_3$Sb$_5$. Here it should be noted that we do not consider an impurity-induced change of the CDW pattern in CsV$_3$Sb$_5$ because the amount of impurity introduced by irradiation is low enough to maintain the double-peak structure of the SC phase in the $P$-$T$ phase diagram\,\cite{Roppongi2023}.

Having established the isotropic $s$-wave SC without sign reversal in the gap function in K/RbV$_3$Sb$_5$, we now discuss ingredients inducing the difference in gap anisotropy between K/RbV$_3$Sb$_5$ and CsV$_3$Sb$_5$. One possibility is the sample quality of the pristine samples since $\rho_0$ is lowest in CsV$_3$Sb$_5$. Indeed, while $\rho_0$ of CsV$_3$Sb$_5$ in Ref.\,\cite{Roppongi2023} is as small as 0.4\,$\mu \Omega$cm, $\rho_0$ of KV$_3$Sb$_5$ and RbV$_3$Sb$_5$ in this study is 3.6 and 2.9\,$\mu \Omega$cm. However, although the 1.3\,C/cm$^2$ electron-irradiated CsV$_3$Sb$_5$ shows $\rho_0=5.9$\,$\mu \Omega$cm\,\cite{Roppongi2023}, which is higher than the values in the pristine K/RbV$_3$Sb$_5$ samples, this sample still exhibits a much smaller value $\Delta_0/k_{\rm B}T_{\rm c}\approx 0.65$ showing stronger anisotropy than in the pristine K/RbV$_3$Sb$_5$ samples. Therefore, the impurity effects cannot wholly explain the isotropic gap structure in K/RbV$_3$Sb$_5$.

Alternatively, we focus on the reconstructions of the band structures in $A$V$_3$Sb$_5$ induced by the distinct $2\times2\times2$ BO states. We consider the $p$-type conduction band composed of the $d_{xz}$-orbitals. In the original kagome lattice model in the absence of the BO, the $p$-type band possesses three vHS points at $M$-points of the original Brillouin zone (BZ). The energy of the vHS points is slightly below the Fermi level; $E_{\rm vHS}\approx -0.06$\,eV for KV$_3$Sb$_5$, $E_{\rm vHS}\approx -0.08$\,eV for RbV$_3$Sb$_5$, and $E_{\rm vHS}\approx -0.1$\,eV for CsV$_3$Sb$_5$. The $p$-type band near the vHS points with large $d_{xz}$-orbital DOS plays an essential role in the emergence of CDW orders (such as the BO and the loop-current order) as well as the SC \cite{Tazai2022,Tazai2023}.

\begin{figure}[t]
    \includegraphics[width=\linewidth]{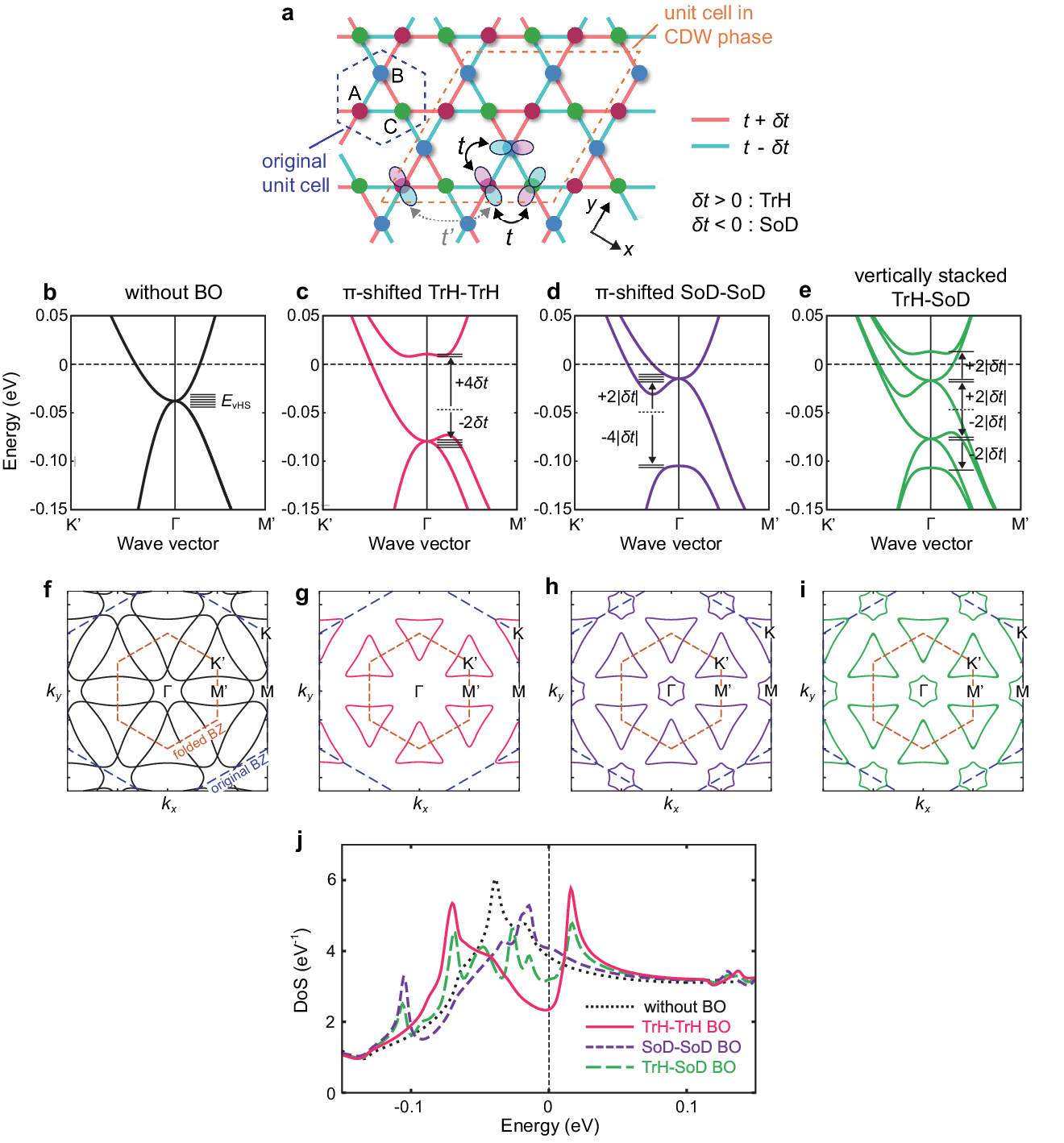}
    \caption{{\bf Band structures in the bond-order states calculated by \textit{d}$_{xz}$-orbital kagome lattice model.}  {\bf a}, TrH BO ($\delta t>0$) and SoD BO ($\delta t<0$) in the two-dimensional $d_{xz}$-orbital kagome lattice model. Hopping integrals are shown. {\bf b}-{\bf e}, Band structure at the $k_z=0$ plane without BO ({\bf b}), and with TrH-TrH BO ({\bf c}), SoD-SoD BO ({\bf d}), and TrH-SoD BO ({\bf e}). Here, we put $t^\perp\rightarrow0$ to simplify the discussion. We set $|\delta t|=0.015$\,eV in {\bf c}-{\bf e}. {\bf f}-{\bf i}, Fermi surfaces without BO ({\bf f}), and with $\pi$-shifted TrH-TrH BO ({\bf g}), $\pi$-shifted SoD-SoD BO ({\bf h}), and vertically stacked TrH-SoD BO ({\bf i}). {\bf j}, $d_{xz}$-orbital DOS in the TrH-TrH BO, SoD-SoD BO, TrH-SoD BO states for $|\delta t|=0.015$\,eV.}
    \label{F4}
\end{figure}

Figure\,\ref{F4}{\bf a} shows the $2\times2$ BO in the $d_{xz}$-orbital kagome lattice model with hopping modulations $\delta t\ne0$: the TrH ($\delta t>0$) and the SoD ($\delta t<0$) BO states. In these BO states, three vHS points are moved to the same $\Gamma$-point in the folded BZ, forming a newly reconstructed band structure made of $d_{xz}$-orbitals. Below, we analyze a three-dimensional $d_{xz}$-orbital model with the kagome lattice given in Fig.\,\ref{F4}{\bf a} for simplicity. The in-plane nearest-neighbor hopping integral is $t=-0.5$\,eV, and the inter-layer hopping integral $t^\perp$ is much smaller than $0.01 |t|$ in magnitude. Hereafter, we set $t^\perp\rightarrow0$ to simplify the discussion. Figure\,\ref{F4}{\bf b} shows the band structure at the $k_z=0$ plane without the BO, while Figs.\,\ref{F4}{\bf c}-{\bf e} show the band structure at the $k_z=0$ in the $\pi$-shifted TrH-TrH, $\pi$-shifted SoD-SoD, and vertically stacked TrH-SoD BO states, respectively. (Here, we set the BO parameter $|\delta t|=0.015$\,eV for each layer.) The corresponding Fermi surfaces are shown in Figs.\,\ref{F4}{\bf f}-{\bf i}, respectively. In the absence of the BO, six vHS energy levels are almost degenerated ($E_{\rm vHS}\approx-0.05$\,eV), composing two small $d_{xz}$-orbital Fermi pockets with large DOS, as shown in Fig.\,\ref{F4}{\bf f}. In the TrH-TrH ($\delta t>0$) and the SoD-SoD ($\delta t<0$) BO states, nearly sixfold degenerated vHS energy levels split into fourfold states at $E_{\rm vHS}-2\delta t$ and twofold states at $E_{\rm vHS}+4\delta t$. In the TrH-TrH BO state, the hybridization gap appears around $\Gamma$ point for $4|\delta t| \gtrsim |E_{\rm vHS}|$, so the two small Fermi pockets disappear, as shown in Fig.\,\ref{F4}{\bf g}. In the TrH-SoD BO state shown in Fig.\,\ref{F4}{\bf e}, six vHS energy levels splits into two single states at $E_{\rm vHS}\pm 4\delta t$ and two twofold states at $E_{\rm vHS}\pm2\delta t$. In this case, single small Fermi pocket survives even for $4|\delta t| \gtrsim |E_{\rm vHS}|$, as shown in Fig.\,\ref{F4}{\bf i}. We note that this characteristic split of vHSs is also observed in the first-principles calculations shown in Supplementary Information. However, it should be noted that folded Fermi pockets are generally difficult to observe in the photoemission spectroscopy due to their intrinsically weak intensity.

Figure\,\ref{F4}{\bf j} compares the $d_{xz}$-orbital DOS in the three (TrH-TrH, SoD-SoD, TrH-SoD) $2\times2\times2$ BO states. Here, we analyze the three-dimensional ($d_{xz}$, $d_{yz}$)-orbital kagome lattice model introduced in Ref.\,\cite{Tazai-EMCHA} to obtain realistic DOS. In the TrH-TrH BO state, the DOS at the Fermi level is reduced by forming the hybridization gap around $\Gamma$-point, while the DOS is essentially unchanged in the SoD-SoD phase. In the TrH-SoD BO state, the DOS at the Fermi level is also reduced by disappearing one of two Fermi pockets around $\Gamma$-point. Therefore, we theoretically revealed that the emergent $d_{xz}$-orbital band structure strongly depends on the three-dimensional BO structure. Note that the $d_{xz}$-orbital DOS at the Fermi level is about 25\% of the total $d$+$p$-orbital DOS according to the first-principles study for $A$V$_3$Sb$_5$. 

Considering the reported differences in the CDW patterns (the TrH-TrH state in K/RbV$_3$Sb$_5$ and the TrH-SoD state in CsV$_3$Sb$_5$)\,\cite{Kang2023,Kautzsch2023,Kato2022,Frassineti2023,Ortiz2021PRX,Li2022PRR,Hu2022}, the small Fermi pockets around the folded $\Gamma$-point composed of the three vHS states survive only in CsV$_3$Sb$_5$ at low temperatures. Note that the above arguments are still valid even if a 2$\times$2$\times$4 BO state is realized in CsV$_3$Sb$_5$ because the presence of the SoD layer is critical to maintain the vHS states around the folded $\Gamma$-point. Theoretically, the $d_{xz}$-orbital vHS states play an essential role in the quantum fluctuations and SC\,\cite{Tazai2022,Tazai2023}. Therefore, although the stacking pattern of BO in our samples is not experimentally confirmed, we expect substantial BO fluctuations derived by the $d_{xz}$-orbital vHSs and associated electron correlations only in CsV$_3$Sb$_5$, which induce an anisotropic gap structure in CsV$_3$Sb$_5$ in contrast to an isotropic gap in K/RbV$_3$Sb$_5$. Furthermore, $T_{\rm c}$ of CsV$_3$Sb$_5$ is more than two times higher than that of K/RbV$_3$Sb$_5$ (Fig.\,\ref{F1}{\bf g}) even though the values of Sommerfeld coefficient $\gamma$ indicate comparable DOS among $A$V$_3$Sb$_5$, as depicted in Fig.\,\ref{F1}{\bf h}, evidencing the critical role of the BO fluctuations to enhance $T_{\rm c}$ in kagome superconductors. Our conclusion implies that the double-peak structure of $T_{\rm c}$ in the $P$-$T$ and hole-doped phase diagrams of CsV$_3$Sb$_5$ may be related to a change in the CDW patterns and associated BO fluctuations across the peak within the BO phase.

Here, we should mention a theoretical study of impurity effects on kagome materials suggesting that sign-changing gap symmetries, including $d$-wave, and chiral $d$-wave states are robust against nonmagnetic disorder because of the sublattice polarization of Fermi surfaces\,\cite{Holbaek2023}. However, this argument is valid when defects on only the V sites are taken into account, and, in other words, defects in the alkali and Sb sites may destroy the sign-changing SC state. Indeed, we can expect a large number of defects on the alkali and Sb sites induced by electron irradiation (see Supplementary Information), and the observed $T_{\rm c}$ reduction in CsV$_3$Sb$_5$\,\cite{Roppongi2023} evidences a finite pair-breaking effect of the nonmagnetic impurities. Moreover, the gap averaging effect of the impurity scatterings observed in CsV$_3$Sb$_5$ cannot be explained by the symmetry-protected gap structures. Therefore, our discussion about the gap symmetry in $A$V$_3$Sb$_5$ is still valid even with the orbital-polarized Fermi surfaces.

It has also been proposed that the vHS derived from the $d_{xy}$ orbital may contribute to CDW formation\,\cite{Denner2021, Barman2024}. While the strong electron correlation of the $d_{xz}$ orbital, which gives the largest DOS near the Fermi level, plays the dominant role, the $d_{xy}$ orbital may also assist the CDW cooperatively via intersite Coulomb or electron-phonon interactions.


In conclusion, we have studied the impurity effects on the SC gap structure and transition temperature in K/RbV$_3$Sb$_5$ from the penetration depth measurements and electron irradiation. By combining our results and the previous report on CsV$_3$Sb$_5$, we provide a comprehensive understanding of the SC gap structure in $A$V$_3$Sb$_5$ ($A$ = K, Rb, Cs): quite anisotropic gap in CsV$_3$Sb$_5$ and isotropic gap in K/RbV$_3$Sb$_5$. Taking into account the differences in the band structures, it turns out that the vHSs essential for BO fluctuations depend sensitively on the patterns of CDW. Significantly developed BO fluctuations in CsV$_3$Sb$_5$ with the SoD pattern play a crucial role in inducing the strong SC gap anisotropy and the relatively high $T_{\rm c}$. Furthermore, $T_{\rm c}$ and $T_{\rm CDW}$ clearly show opposite trends against disorder, suggesting intense competition between SC and CDW phases. Our systematic studies on $A$V$_3$Sb$_5$ shed new light on the interplay between unconventional $s$-wave SC and CDW orders in kagome materials and hopefully promote further exploration of novel physics originating from an interplay of exotic orders in kagome systems.

\section*{Methods}
\subsection*{Single crystal growth}
High-quality single crystals of K/RbV$_3$Sb$_5$ were grown by a modified self flux method using K ingot (Alfa, 99.95\%), V powder (Sigma, 99.9\%), and Sb shot (Alfa, 99.999\%) for KV$_3$Sb$_5$ and Rb ingot (Alfa, 99.75\%), V powder (Sigma, 99.9\%,) and As shot (Alfa, 99.999\%) for RbV$_3$Sb$_5$, respectively. The mixture was placed in an alumina crucible, which was then sealed in a quartz ampoule under high vacuum. The sealed ampoule was heated to 1000 ${}^\circ$C, soaked at this temperature for 24 hours, and subsequently cooled down. Single crystals were then mechanically extracted from the flux. All preparation steps, except for the sealing and heating processes, were carried out in an argon glovebox.

\subsection*{Magnetic penetration depth measurement}
The temperature dependence of the penetration depth was measured by the tunnel diode oscillator (TDO) technique at 13.8 MHz in a dilution refrigerator. The sample was put at the center of a coil forming the TDO circuit, and the relative change of the penetration depth $\Delta\lambda (T)=\lambda (T)-\lambda (0)$ is directly obtained from the shift of the resonant frequency of the TDO circuit $\Delta f(T)$ with the relation $\Delta\lambda (T) =G \Delta f(T)=f(T)-f(0)$, where the geometrical constant $G$ is determined by the shape of the sample. Here, the magnetic field induced by the coil is perpendicular to the kagome planes with a magnitude of the order of $\mu$T, which is much lower than the lower critical filed of the order of mT, confirming the Meissner state of our samples.

\subsection*{Electrical resistivity measurement}
The electrical resistivity was measured by the four-terminal method with dc current applied within the $ab$ plane in a $^3$He refrigerator for $T<5$\,K and in a home-made probe for $T>4.2$\,K. The delta-mode of Keithley model 6221 and 2182A was used to eliminate the offset of dc resistivity. We confirmed that the applied current was low enough that the Joule heating can be ignored. 

\subsection*{Electron irradiation}
Electron irradiation with the incident energy of 2.5 MeV was performed on the SIRIUS Pelletron accelerator in Laboratoire des Solides Irradies (LSI) at Ecole Polytechnique. Here, energy transfer from the irradiated electrons to the lattice exceeds threshold energy for the formation of vacancy-interstitial Frenkel pairs, which act as point defects. We performed irradiation around 20 K to prevent the defect migration and agglomeration. Although partial annealing of the introduced defects occurs on warming to the room temperature, uniform point defects are kept due to the lower migration energy. Electron irradiation has no pressure or doping effect since it does not change the lattice constants and carrier density.

\subsection*{Theoretical calculations}
In this study, we analyze an effective Hamiltonian model based on an extended unit cell with 24 sites\,\cite{Tazai-EMCHA}. Each unit cell consists of two layers, with 12 V sites per layer. The Hamiltonian is given by:

\begin{align*}
  H = \sum_{i,j, l,m, \sigma} t_{i,j}^{l,m} c^{\dagger}_{i,l,\sigma} c_{j,m,\sigma} 
  + \sum_{\left< i,j \right>, l,m, \sigma} \delta t_{i,j} c^{\dagger}_{i,xz,\sigma} c_{j,xz,\sigma}.
\end{align*}

Here, $c^{\dagger}_{i,l,\sigma}$ is the creation operator for a spin-$\sigma$ electron in orbital $l$ (either $d_{xz}$ or $d_{yz}$) on the $i$-th site ($i = 1$ to $24$). The hopping parameters $t_{i,j}^{l,m}$ are chosen to approximately reproduce the three-dimensional Fermi surface obtained from first-principles calculations. Specifically, the onsite energy of the $d_{yz}$ orbital is set to $2.3$. The nearest-neighbor hoppings are set as follows: $t = -0.5$ for intra-orbital $d_{xz}$ hopping, $t_{yz} = -1$ for $d_{yz}$, and $t_{xz \textrm{-} yz} = \pm 0.05$ for inter-orbital $d_{xz}\textrm{-}d_{yz}$ hopping. The hopping across two sites within the $d_{xz}$ orbital is set to $t' = -0.08$ (see Fig.\,\ref{F4}{\bf a}). We set interlayer hopping of the intra-$d_{yz}$ orbital $t^{\perp}_{yz,yz} = 0.02$ for the site directly above and those belonging to the same triangle. The interlayer hopping for the $d_{xz}$ orbital is neglected due to its small magnitude. The unit for all hopping parameters is eV.

The bond order is introduced as a modulation of the nearest-neighbor hopping in the $d_{xz}$ orbital, $\delta t^b_{i,j} = \pm \delta t$, with the modulation pattern shown in Fig.\,\ref{F4}{\bf a}.

For Fig.\,\ref{F4}, we used a $240 \times 240 \times 120$ $k$-mesh. The electron number was set to $N = 17.8$. For Figs.\,\ref{F4}{\bf b}-{\bf i}, in order to clarify the changes in the band structure and Fermi surface, we used a single-orbital model with only the $d_{xz}$ orbital and set $N = 11.6$.

\section*{Data availability}
The data that support the findings of this study are available from the corresponding authors upon reasonable request.

\section*{Code availability}
The codes used for the numerical calculations are available from the corresponding authors upon reasonable request. The commercially available WIEN2k software was also used for the first principles calculations presented in the Supplementary Information.

\section*{Acknowledgments}
This work was supported by Grants-in-Aid for Scientific Research (KAKENHI) (Nos.\ JP24K17007, JP24H01646, JP23H00089, JP22H00105), and Grant-in-Aid for Scientific Research for Transformative Research Areas (A) ``Correlation Design Science'' (No.\ JP25H01248) from Japan Society for the Promotion of Science (JSPS). Electron irradiation was conducted at the SIRIUS accelerator facility at {\' E}cole Polytechnique (Palaiseau, France) and was supported by EMIR\&A French network (FR CNRS 3618) (proposal No.\ 22-8950). S.D.W. and A.C.S. gratefully acknowledge support via the UC Santa Barbara NSF Quantum Foundry funded via the Q-AMASE-i program under award DMR-1906325.

\bibliographystyle{naturemag}
\bibliography{AV3Sb5}

\clearpage
    
\end{document}